\documentclass[twocolumn]{emulateapj}
\usepackage{graphicx}
\usepackage{color}
\usepackage{soul}

\usepackage{float}
\usepackage[pdftex,
            bookmarks=true,
            breaklinks,
            pagebackref=true]{hyperref}         
            
\usepackage{multirow}
\usepackage{threeparttable}

\usepackage{savesym}
\savesymbol{tablenum}
\usepackage{siunitx}
\restoresymbol{SIX}{tablenum}

\shorttitle{Modal Noise Mitigation}
\shortauthors{Petersburg et al.}

\begin{document}

\title{Modal Noise Mitigation through Fiber Agitation for Fiber-fed Radial Velocity Spectrographs}

\author{Ryan R. Petersburg, Tyler M. McCracken, Dominic Eggerman, Colby A. Jurgenson, David Sawyer, Andrew E. Szymkowiak, Debra A. Fischer}
\affil{Department of Astronomy, Yale University, 52 Hillhouse Avenue, New Haven, CT 06511, USA; ryan.petersburg@yale.edu}

\begin{abstract}

Optical fiber modal noise is a limiting factor for high precision spectroscopy signal-to-noise in the near-infrared and visible. Unabated, especially when using highly coherent light sources for wavelength calibration, modal noise can induce radial velocity (RV) errors that hinder the discovery of low-mass (and potentially Earth-like) planets. Previous research in this field has found sufficient modal noise mitigation through the use of an integrating sphere, but this requires extremely bright light sources, a luxury not necessarily afforded by the next generation of high-resolution optical spectrographs. Otherwise, mechanical agitation, which ``mixes'' the fiber's modal patterns and allows the noise to be averaged over minutes-long exposures, provides some noise reduction but the exact mechanism behind improvement in signal-to-noise and RV drift has not been fully explored or optimized by the community. Therefore, we have filled out the parameter space of modal noise agitation techniques in order to better understand agitation's contribution to mitigating modal noise and to discover a better method for agitating fibers. We find that modal noise is best suppressed by the quasi-chaotic motion of two high-amplitude agitators oscillating with varying phase for fibers with large core diameters and low azimuthal symmetry. This work has subsequently influenced the design of a fiber agitator, to be installed with the EXtreme PREcision Spectrograph, that we estimate will reduce modal-noise-induced RV error to less than \SI{3.2}{\centi\meter\per\second}.

\end{abstract}

\section{Introduction}
\label{sec:intro}

Radial velocity (RV) exoplanet detection has continuously been on the path toward higher precision to enable the detection of less massive and longer period planets. The current goal of RV spectroscopy is \SI{10}{\centi\meter\per\second} precision, a factor of 10 better than current state-of-the-art RV spectroscopy, thereby allowing the discovery of Earth-like planets orbiting G and K stars in their respective habitable zones \citep{Fischer2016}. The next-generation of visible-band RV spectrographs---including but not limited to the EXtreme PREcision Spectrograph (EXPRES; \citet{Jurgenson2016}), ESPRESSO \citep{Megevand2012}, NEID \citep{Schwab2016}, and the Keck Planet Finder \citep{Gibson2016}---require precision engineering and extreme stability to reach this goal.

Fiber coupling the spectrograph to the telescope has become an essential and standard method for planet hunting spectrographs. Separating the spectrograph from the telescope by a fiber tens of meters long enables the spectrograph to be located in a controlled environment, isolating it from vibrational and thermal noise. Linking the telescope to the spectrograph via fiber also leverages the spatial scrambling properties inherent to fibers that, for the most part, decouple input variations from the output producing a stable illumination of the spectrograph optics \citep{Hunter1992}. This effect has been amplified through the use of double scramblers \citep{Halverson2015a, Spronck2015} and non-circular fiber geometries \citep{Chazelas2010, Spronck2012a, Plavchan2013}.

Optical fibers also transmit light from calibration sources, such as wavelength calibrators and broadband flat-field sources, to the spectrograph. Laser frequency combs, especially the  astrocomb \citep{Probst2014} recently deployed at HARPS and soon at EXPRES, produce thousands of ultra-narrow, evenly spaced emission lines over a wide frequency range. When these highly coherent lines propagate through a multi-mode fiber, they create a source of noise that limits the signal-to-noise ratio (S/N) of the instrument and potentially induces false RV signals in the data. This noise is caused by interference between the finite number of electromagnetic modes that can propagate along a multi-mode fiber, and therefore the term \textit{modal noise} has been coined for this effect \citep{Epworth1978}.

Some next-generation RV spectrographs---e.g. iLocater \citep{Crepp2016} and MINERVA-red \citep{Blake2015}---have moved to a completely single-mode fiber architecture to help alleviate these complications. As apparent in their name, single-mode fibers only propagate a single spatial mode and should be free from any modal noise. Due to the small core size of single-mode fibers, however, coupling light from the telescope into these fibers is challenging and requires robust adaptive optics not currently available in the visible. Single-mode fibers also have a limited bandwidth---approximately 100--\SI{200}{\nano\meter} in the visible and 400--\SI{600}{\nano\meter} in the near infrared---over which they propagate a single mode. Thus, the fiber architectures for these spectrographs require multiple band-dependent paths or endlessly single-mode photonic crystal fibers (yet untested in RV spectroscopy) for broadband coverage. It is also possible that single-mode fibers are not free from modal noise since they propagate in two polarization modes that have been shown to affect spectrograph performance \citep{Halverson2015}. The study of modal noise reduction methods may still be necessary even regarding the existence of these novel single-mode fiber-fed instruments.

In this paper, we attempt to discern the optimal strategy for reducing modal noise using mechanical agitation on multi-mode fibers propagating coherent visible light. We begin by defining modal noise and exploring how previous experiments have mitigated it through static and dynamic methods (Section \ref{sec:modal_noise_intro}). We then describe our own methods of fiber agitation (Section \ref{sec:experimental_setup}) and discuss results from using these methods on fibers of varying cross-sectional shapes and coupling permutations (Section \ref{sec:results}). Finally, we relate these results to limits in RV precision (Section \ref{sec:rv_precision}), test how our agitation methods affect focal ratio degradation (FRD; Section {\ref{sec:frd}}), and discuss how these results should be applied to next-generation RV spectrographs (Section \ref{sec:conclusions}). The work in this paper was conducted to influence design decisions for EXPRES.

\section{Optical Fiber Modal Noise}
\label{sec:modal_noise_intro}

Light propagates through an optical fiber in an integer number of electromagnetic modes. The exact calculation for this value is nontrivial since it depends on the instantaneous fiber geometry, injection parameters, and many other variables. The maximum number of modes for a step-index circular cross-section fiber (propagating a relatively large number of modes) with a monochromatic light source is approximately
\begin{equation}
M_{circ} \approx \frac{4}{\pi ^2} V^2 = \frac{4}{\pi ^2} \Bigg( \frac{2 \pi r \mathrm{NA}}{\lambda} \Bigg) ^2.
\label{eq:max_modes}
\end{equation}
$V$ is the normalized frequency of the fiber rewritten in terms of $\mathrm{NA} = \sqrt{n_\mathrm{core}^2 - n_\mathrm{clad}^2}$, the numerical aperture of the fiber determined by the core ($n_\mathrm{core}$) and cladding ($n_\mathrm{clad}$) indices of refraction, the core radius $r$, and the wavelength $\lambda$ of propagated light. This approximation is more difficult for a rectangular fiber, but \citet{Nikitin2011} shows empirically using electromagnetic and geometrical arguments that
\begin{equation}
\frac{M_\mathrm{rect}}{M_\mathrm{circ}} \approx 2 \frac{ab}{\pi r^2}
\label{eq:prop_modes}
\end{equation}
where $a$ and $b$ are the side lengths of the rectangular cross-section. Notice that $ab$ and $\pi r^2$ give the areas for a rectangle and circle respectively. From this, we will assert more generally, with some rearrangement of Equation (\ref{eq:max_modes}), that
\begin{equation}
M_{s} \approx \frac{16}{\pi} C_{s} A \Bigg( \frac{\mathrm{NA}}{\lambda} \Bigg) ^2
\label{eq:mode_area}
\end{equation}
where $A$ is the cross-sectional area of the fiber and $C_{s}$ is a constant coefficient dependent on fiber cross-sectional shape such that $C_\mathrm{circ} = 1$ and $C_\mathrm{rect} = 2$. $C_{s}$ is so far unknown for more complicated geometries, but we assume that $C_{s} \sim 1$.

\begin{figure}
\centering
	\includegraphics[width=\columnwidth]{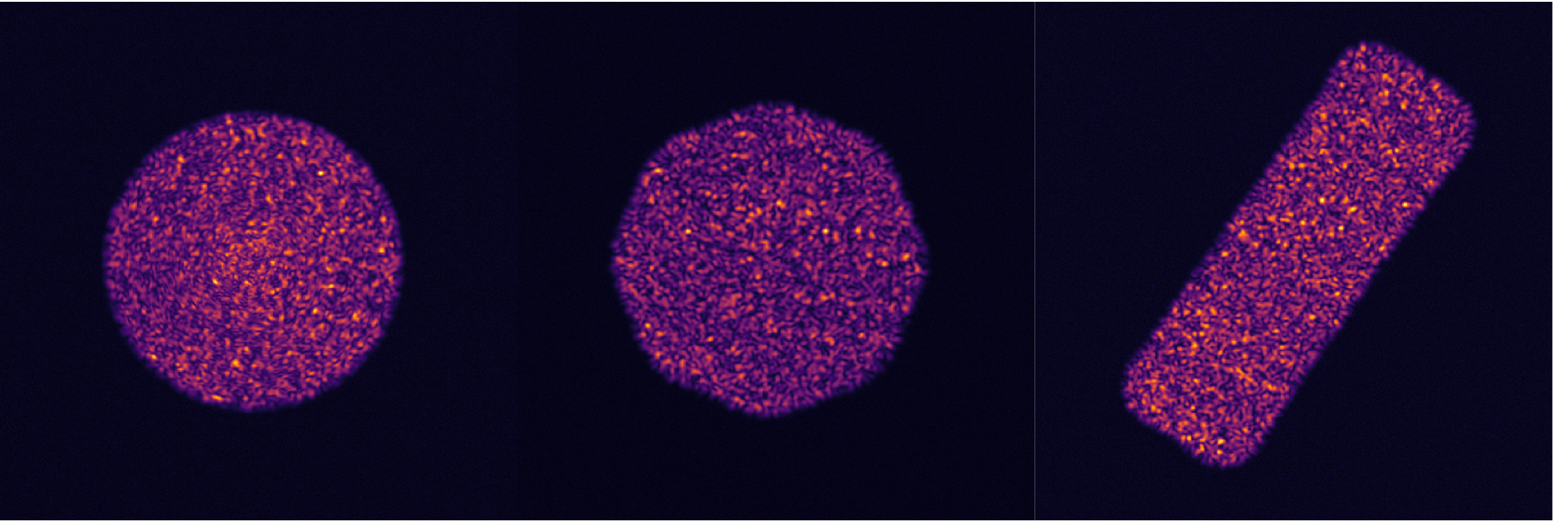}
	\caption{Examples of unmitigated modal noise for a \SI{200}{\micro\meter} circular (left), \SI{200}{\micro\meter} octagonal (middle), and \SI{100x300}{\micro\meter} rectangular (right) optical fiber. All three fibers shown here have approximately the same cross-sectional area, meaning that they each are propagating about the same number of electromagnetic modes (see Equation (\ref{eq:mode_area})). Brightness in this image is scaled by (photon count)$^{0.6}$ for better presentation of the range of speckle brightness. The contrast of the speckles is therefore \textit{worse} than what is shown here.}
\label{fig:fiber_example}
\end{figure}

When coherent light is propagated through a multi-mode fiber, a high contrast speckle pattern known as modal noise is produced at the output for both the near field (fiber face projected onto a detector, Figure \ref{fig:fiber_example}) and far field. Modal noise is an inherent property of all multi-mode fibers regardless of the cross-sectional core shape \citep{Sablowski2016}. It arises from light coupling from mode-to-mode as it propagates through the fiber, causing slight variances in the path length traveled and producing the observed interference pattern. For RV spectrographs, this causes two problems: (1) it limits the maximum S/N and (2) systematic variations in the speckle pattern will mask themselves as minute shifts on the focal plane causing errant RV signatures.

\subsection{Limit on S/N}

Due to its high contrast, modal noise can severely decrease the S/N of an RV spectrograph \citep{Epworth1978, Baudrand2001, Lemke2011, Iuzzolino2014}. For a fiber without spatial filtering or a slit, the magnitude of this noise is proportional to $\sqrt{M_s}$ \citep{Goodman1981}. Therefore, increasing the size of the fiber, increasing the numerical aperture of the fiber (or decreasing the injected focal ratio), and decreasing the wavelength of injected light should increase the S/N due to modal noise. It also appears that changing the fiber core shape could affect the S/N.

Experimental results of these conditions have been well documented. S/N due to modal noise has been shown empirically to
\begin{enumerate}
\item increase with larger fiber core cross-sectional area \citep{Lemke2010, Sablowski2016},
\item decrease with larger focal ratios \citep{Baudrand2001, Sablowski2016},
\item decrease with longer wavelength of injected light \citep{Baudrand2001},
\item slightly increase with more static bends in the fiber (changing the NA; \citet{Imai1979}),
\item remain the same for differing fiber lengths greater than a few meters \citep{Baudrand2001}, and
\item improve for non-circular fibers over circular fibers \citep{Sablowski2016, Sturmer2016}.
\end{enumerate}
All of these results follow exactly from Equation (\ref{eq:mode_area}) and implies that fibers with non-circular geometries have larger $C_{s}$.

\subsection{Systematic variations}
\label{subsec:sys_var}

Since the resultant speckle pattern is dependent on dynamic optical properties of the fiber, modal noise can induce false RV's on the spectrograph \citep{Mahadevan2014}. The spectrograph input is directly imaged onto the detector, so any spatial variation in intensity of the injected light will change the apparent line spread function in the spectra and cause errors in the assigned RVs. If these drifts have some regular period, modal noise could even cause errant planet signatures.

As is clear in Equation (\ref{eq:mode_area}), modal noise is heavily wavelength dependent. When the center wavelength of a coherent light source changes, the speckle pattern subsequently shifts.  Speckle patterns can be visually distinguished when the injected wavelength of light changes by at least \SI{8}{\pico\meter} at \SI{1500}{\nano\meter} \citep{Redding2013}. Next-generation RV spectrographs are using laser frequency combs with $10^{-11}$ stability \citep{Probst2014}, or less than $10^{-4}$ \SI{}{\pico\meter} stability at \SI{1500}{\nano\meter}, rendering speckle drift due to wavelength drift effectively irrelevant. However, since the speckle pattern is smoothly wavelength dependent, the resultant spectral line spread function of a frequency comb is correlated between neighboring lines, meaning any drift due to modal noise is not necessarily randomly distributed across the spectrum.

The speckle pattern seen at the end of a fiber changes over time most commonly because of \citep{Epworth1978}:
\begin{enumerate}
\item temperature variation,
\item fiber input illumination variation, and
\item fiber movement (bending, twisting, etc.).
\end{enumerate}
These three conditions inevitably pose problems when imaging a spectrum since they are inherent to modern fiber-fed RV spectrographs \citep{Baudrand2001, Mahadevan2014}. There is typically a changing temperature differential between the telescope and the spectrograph, fluctuations in atmospheric density and guiding change the fiber illumination, and the telescope (along with the connected fibers) slowly moves throughout the night.

\subsection{Mitigation Techniques}
\label{subsec:mitigation}

\begin{table*}
\centering
\small
\caption{Previous Study of Dynamic Modal Noise Mitigation Methods}
	\small
	\begin{tabular}{llcc}
		\hline
		References & Method & Frequency & Amplitude \\
		\hline\hline
		\citet{Daino1980} & Loudspeaker & \SI{110}{\hertz} & ``Sufficient'' \\
		\hline
		\citet{Hill1980} & Turbulent Air Stream & ---- & --- \\
		\hline
		\citet{Baudrand2001} & --- & \SI{30}{\hertz} & \SI{1}{\milli\meter} \\
		\hline
		\multirow{2}{*}{\citet{Lemke2011}} & Loudspeaker & 1.5 Hz & --- \\
		 & Loudspeaker & \SI{80}{\hertz} & --- \\
		\hline
		\multirow{3}{*}{\citet{McCoy2012}} & Paint mixer & \SI{60}{\hertz} & --- \\
		 & Hand agitated & 1-\SI{2}{\hertz} & 10-\SI{15}{\centi\meter} \\
		 & Mechanical agitator & 2-\SI{3}{\hertz} & 1-\SI{5}{\centi\meter} \\
		\hline
		\multirow{2}{*}{\citet{Plavchan2013}} & ``Tweeter'' & \SI{100}{\hertz} & \SI{1}{\milli\meter} \\
		 & ``Woofer'' & \SI{1}{\hertz} & \SI{25}{\milli\meter} \\
		\hline
		\multirow{3}{*}{\citet{Mahadevan2014}} & Int. Sph. + Diff. & --- & ---\\
		 & McCoy agitator & 2-\SI{3}{\hertz} & 1-\SI{5}{\centi\meter} \\
		 & Hand agitation & 1-\SI{2}{\hertz} & \SI{10}{\centi\meter} \\
		\hline
		\multirow{2}{*}{\citet{Halverson2014}} & Int. Sph. + Diff. & --- & --- \\
		 & Int. Sph. + Rot. Mirror & --- & --- \\
		\hline		
		\citet{Roy2014} & Rail agitator & 1-\SI{2}{\hertz} & \SI{170}{\milli\meter} \\
		\hline
		\citet{Sablowski2016} & ``Rotating Excenter'' & \SI{2}{\hertz} & \SI{20}{\centi\meter} \\
		\hline
	\end{tabular}
\label{table:previous_studies}
\end{table*}

Modal noise can be mitigated by continuously exacerbating one of the above three dynamic variations, thereby shifting the speckle pattern throughout an appropriately long camera exposure and averaging out the noise. Controlled temperature variation (option 1) is non-ideal because a \SI{1}{\meter} fiber requires approximately $8 ^\circ \mathrm{C}$ amplitude fluctuations to visibly decorrelate the speckle pattern \citep{Redding2013}, and would be impractical to implement. Therefore, RV spectrographs have been left with either varying the illumination (option 2) or shaking the fiber (option 3). As summarized in Table \ref{table:previous_studies}, these modal noise reduction techniques have been discussed by many experiments concerned with RV spectroscopy.

\citet{Mahadevan2014} and \citet{Halverson2014} explore the effectiveness of varying the illumination on the fiber face. Using an integrating sphere, diffuser, and rotating mirror, they show gradual improvements in modal noise reduction due to the addition of further illumination variation. However, the integrating sphere, an integral part of these methods, has a throughput efficiency of approximately $10^{-6}$ and is not feasible to be introduced in the science light optical path. To allow flexible observing programs, particularly science observations bracketed by precision wavelength calibration sources, the modal noise mitigation technique needs to be more efficient.

Otherwise, the majority of these studies use various forms of agitation---including loudspeakers, paint mixers, and air streams---that shake the fiber over time. The variation in frequency and amplitude for these methods is unfortunately quite wide and conclusions are difficult to make. However, there have been slight trends in the results and the discussed assumptions so far are as follows:
\begin{enumerate}
\item The frequency of agitation should be greater than $1/\tau$, where $\tau$ is the exposure time \citep{Baudrand2001}.
\item Noise is more effectively reduced by high-amplitude motion \citep{Lemke2011, McCoy2012}.
\item More oscillations per exposure time (with an upper limit) provide further noise reduction \citep{Lemke2011}.
\item Combining a high-frequency ``tweeter'' with a high-amplitude ``woofer'' reduces noise effectively \citep{Plavchan2013}.
\item Hand agitation is better than any form of mechanical agitation \citep{Lemke2011, McCoy2012, Mahadevan2014, Roy2014}.
\item Non-harmonic or chaotic motion is recommended \citep{Grupp2003} though an exact method has not yet been experimentally tested.
\end{enumerate}
Although this has been good for subjective intuition, the exact mechanisms behind the improvements in S/N and prevention of RV drift due to fiber agitation have not yet been explored.

In the following sections, we fill out the parameter space of fiber agitation methods further than previous studies. We are interested in seeing trends across different agitation amplitudes and frequencies, fiber shapes and sizes, and coupling permutations to make more precise conclusions about the nature of modal noise mitigation through fiber agitation.

\section{Experimental Setup}
\label{sec:experimental_setup}

The number of modes a fiber supports is largely determined by its cross-sectional area (see Equation (\ref{eq:mode_area})). Also, RV spectrograph fiber architectures typically maintain fiber cross-sectional area for consistent \'etendue and low light loss when reimaging between fibers. For these two reasons, we choose fibers with similar cross-sectional areas when testing and characterizing agitation methods across multiple fiber geometries. Table \ref{table:fibers} lists the fibers used in our experiment. Notice that the \SI{200}{\micro\meter} circular, \SI{200}{\micro\meter} octagonal, and \SI{100x300}{\micro\meter} rectangular fiber all support a similar number of modes.

\begin{table}
\centering
\begin{threeparttable}
\small
\caption{Tested Optical Fibers.}
	\small	
	\begin{tabular}{lll}
	\hline
	Shape & Size & Manufacturer \\
	\hline\hline
	Circular & \SI{100}{\micro\meter} & Polymicro \\
	Circular & \SI{200}{\micro\meter} & Polymicro \\
	Octagonal & \SI{100}{\micro\meter} & CeramOptec \\
	Octagonal & \SI{200}{\micro\meter} & CeramOptec \\
	Rectangular & \SI{100x300}{\micro\meter} & CeramOptec \\
	\hline
	\end{tabular}
	\begin{tablenotes}
	\item \textbf{Note.} All Fibers Have $\mathrm{NA} = 0.22$
	\end{tablenotes}
\end{threeparttable}
\label{table:fibers}
\end{table}

\begin{figure}
\centering
	\includegraphics[width=\columnwidth]{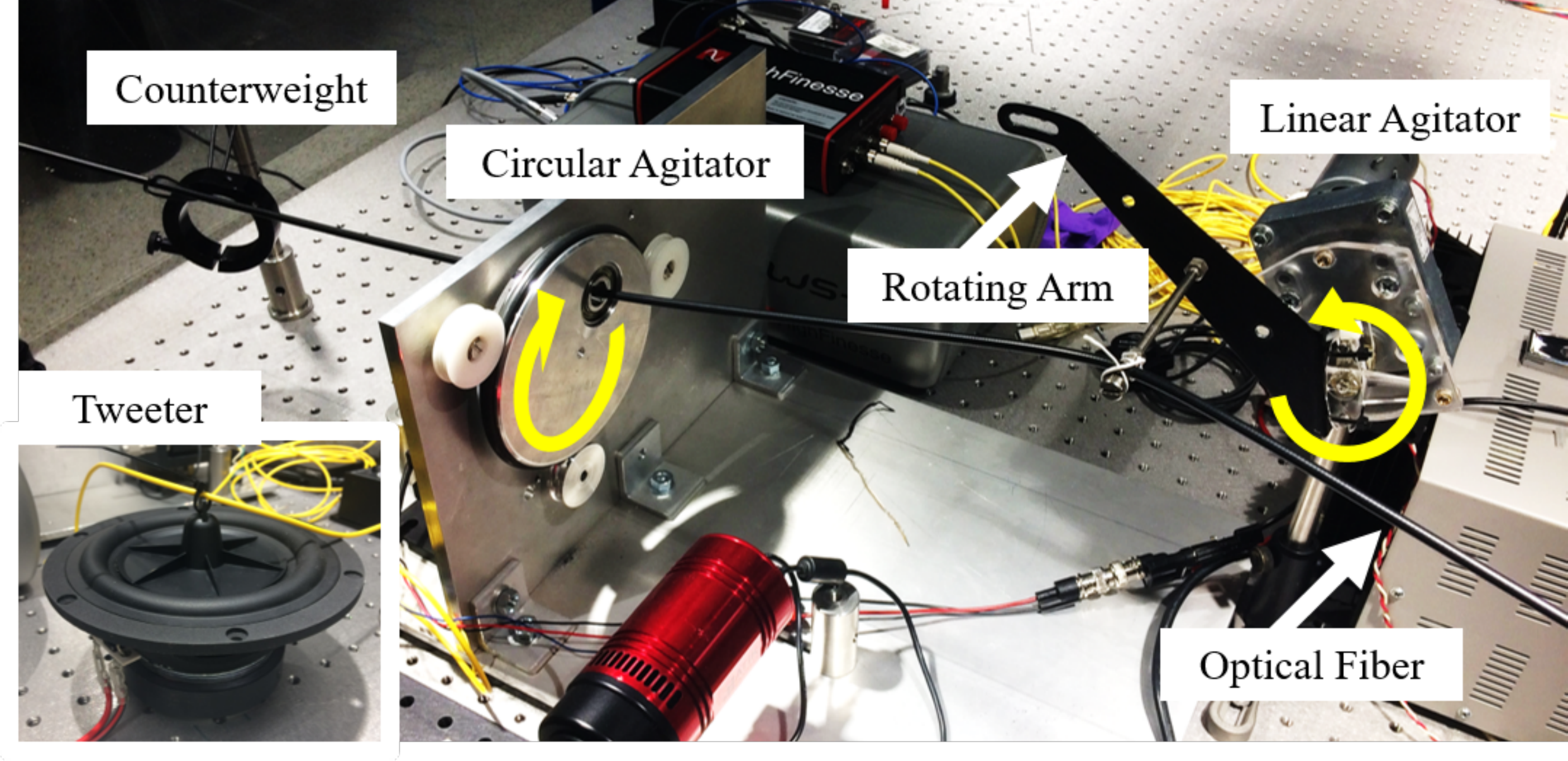}
	\caption{Laboratory images of the linear and circular agitator used in these modal noise tests attached to the \SI{100x300}{\micro\meter} rectangular fiber. The linear agitator rotates a variable-amplitude arm parallel to the fiber while the circular agitator rotates perpendicular to the fiber. A small counterweight keeps a minimal amount of tension in the fiber and prevents it from over-bending or bunching up. The inset in the bottom left contains a PASCO Scientific economy wave driver (``tweeter'') attached to an optical fiber. The tweeter is used to test high-frequency, low-amplitude agitation.}
\label{fig:agitators}
\end{figure}

Two different methods of large-amplitude mechanical agitation are tested (Figure \ref{fig:agitators}): the first produces a linear-type motion in which the fiber is moved up and down, the other is a circular type motion in which the fiber is rotated perpendicular to the direction of propagation. The linear agitator has variable amplitude allowing for 80--\SI{320}{\milli\meter} peak-to-peak amplitude agitation at \SI{80}{\milli\meter} intervals and variable frequency in the range of 0.03--\SI{1.0}{\hertz}. For the circular agitator, the fiber is rotated in a circular path with a set diameter (peak-to-peak amplitude) of \SI{80}{\milli\meter} and a variable frequency from 0.1 to \SI{1.5}{\hertz}. Routing a fiber through both agitators produces what we call ``coupled agitation.''  Frequencies are independently set by adjusting the appropriate DC-motor drive voltage and the amplitude of the linear agitator is set by the position of the lifting arm. A small counterweight is present to keep a minimal amount of tension in the fiber between the agitators and prevent it from folding on itself.

To test high-frequency, low-amplitude agitation, we use a PASCO Scientific ``Economy Wave Driver'' shown in the inset of Figure \ref{fig:agitators}. This device, attached to a sine wave function generator, produces approximately \SI{5}{\milli\meter} amplitude for 10-\SI{30}{\hertz} oscillations. It can be driven at higher frequencies, but the amplitude would not be large enough to produce significant fiber motion. We call this device a ``tweeter'' in homage to \citet{Plavchan2013}.

\begin{figure}
\centering
	\includegraphics[width=\columnwidth]{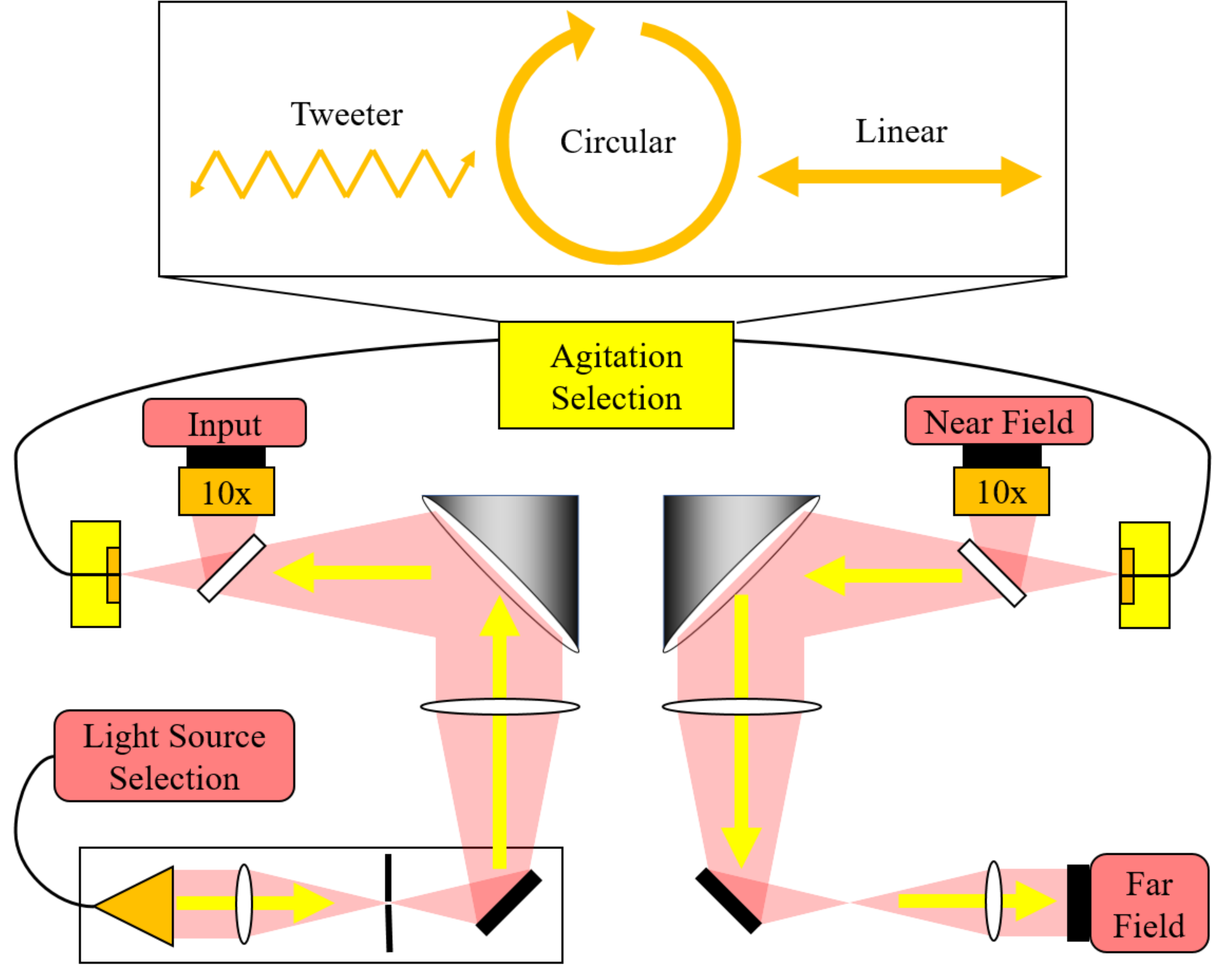}
	\caption{Schematic of the Fiber Characterization Station. Our choice of light source is fed into the station in the bottom left. This light is spatially filtered at focus, collimated (with optional iris for NA selection), and finally injected into the test fiber as a \SI{10}{\micro\meter} spot. The injection face of the test fiber is imaged at $10\times$ magnification by the input camera to allow for precision alignment. Light propagates through the test fiber and our choice of agitator mixes the modes. A cartoon of the three types of agitation (see Figure \ref{fig:agitators}) is presented above the schematic. Light is then ejected from the test fiber and split between the $10\times$ magnified near-field camera and the far-field camera.}
\label{fig:fcs}
\end{figure}

All image data is collected with the Fiber Characterization Station (FCS, Figure \ref{fig:fcs}), a multipurpose device that is able to simultaneously image the input face, near field, and far field of the fiber under test. For these tests, we feed the FCS with either a \SI{652}{\nano\meter} Toptica diode laser (less than \SI{1}{\mega\hertz} linewidth) through a single-mode fiber or a broadband Thorlabs mounted LED centered at approximately \SI{455}{\nano\meter} through a \SI{100}{\micro\meter} circular fiber. Regardless of light source, the FCS focuses light into the fiber as a \SI{10}{\micro\meter} Gaussian spot. Specifications for the FCS cameras are listed in Table \ref{table:cameras}. According to these specifications, our near-field camera has a spatial resolution of \SI{0.3}{\micro\meter}. However, by subtracting ambient calibration images, strictly thresholding to remove background counts, and comparing the unweighted and weighted centroids of each fiber image (thus removing camera drift), we have yielded fiber-centroiding precision to about \SI{0.01}{\micro\meter}.

\begin{table}
\centering
\caption{Fiber Characterization Station Imaging Specifications}
	\begin{tabular}{llcc}
	\hline
	Name & Camera & Pixel Size & Magnification \\
	\hline \hline
	Input & Atik 450 & \SI{3.45}{\micro\meter} & 10 \\
	\hline
	Near Field & Atik 450 & \SI{3.45}{\micro\meter} & 10 \\
	\hline
	Far Field & Atik 383L+ & \SI{5.4}{\micro\meter} & N/A \\
	\hline	
	\end{tabular}
\label{table:cameras}
\end{table}

Image exposure times are set according to the frequency of agitation such that each exposure lasts exactly one period of rotation. For example, if an agitator is set to rotate at \SI{0.5}{\hertz}, each image will be exposed for \SI{2.0}{\second}. We also scale the intensity of our light source to the set exposure time to minimize the relative effect of read-noise on short exposure images. \citet{Baudrand2001} and \citet{Lemke2011}, as outlined in Section \ref{subsec:mitigation}, find that more rotations within the same exposure time improve modal noise. We rather want to see if frequency improves modal noise \textit{per rotation}. Therefore, we control for ``number of rotations'' by confirming exactly one rotation is being recorded.

Each data set is comprised of 10 exposures for each of the following cases:
\begin{enumerate}
\item The fiber being actively agitated.
\item The fiber routed through the agitator but without agitation (``unagitated'').
\item The unagitated fiber illuminated with a broadband LED.
\end{enumerate}
Multiple images are taken (1) to reduce statistical errors in our S/N calculations potentially caused by camera noise or small inconsistencies with our agitators and (2) to observe the effects of agitation at longer exposure times by coadding multiple single-rotation images together. The LED source acts as a control for our S/N and centroiding noise floor, since it is relatively incoherent and thus modal noise is suppressed when using wavelength-integrating cameras, and the unagitated fiber acts as a worst-case scenario.

\begin{figure}
\centering
	\includegraphics[width=\columnwidth]{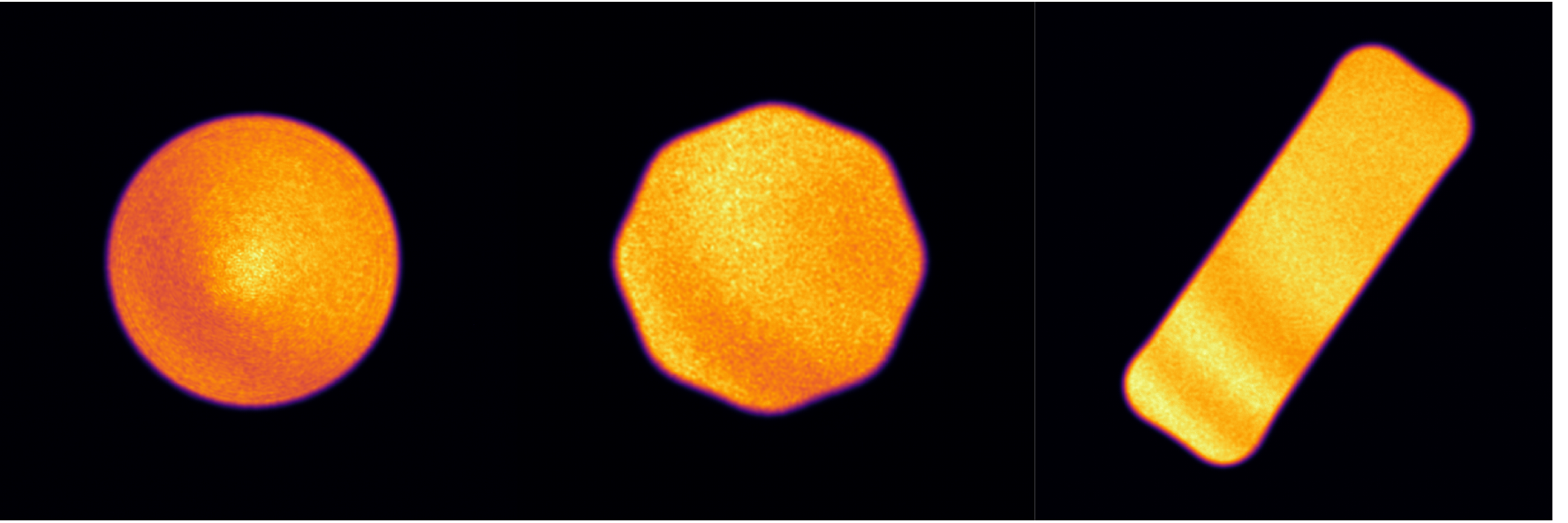}
	\caption{10-rotation, coupled agitation images using the same three fibers shown in Figure \ref{fig:fiber_example}.  The diffraction pattern is clearly seen, but otherwise, the presence of speckles in these images is significantly diminished. Brightness is directly scaled with photon count in these images.}
\label{fig:fiber_improved}
\end{figure}

We quantify modal noise using the S/N of light within the fiber face of the near-field image calculated as
\begin{equation}
\mathrm{S/N} = \frac{\mathrm{median}(I_\mathrm{filt})}{\mathrm{stdev}(I_0 - I_\mathrm{filt})}
\label{eq:snr}
\end{equation}
where $I_0$, the original raw image, is heavily median filtered to produce $I_\mathrm{filt}$. The typical S/N (where the ``signal'' is assumed to be a top-hat function across the fiber face) could not be used due to slight intensity-varying diffraction effects across the near-field image (see Figure \ref{fig:fiber_improved}). The contrast from these diffraction effects is approximately 0.2 and we presume they are caused by the thin-film beamsplitter used on the FCS and not from the fiber itself. Subtracting $I_\mathrm{filt}$ from $I_0$ therefore produces a noise pattern that reflects the modal noise and not these large-scale diffraction-caused variances.

We use a circular 51 pixel median filter rather than a low-order polynomial or Gaussian fit because these latter functions were too smooth to provide a good fit to the raw fiber images. The size of the filter kernel was chosen such that speckles on unagitated images are sufficiently filtered without removing structure from the edges of the fibers. The numerator in Equation (\ref{eq:snr}) is calculated as the median (rather than the mean, for example) of $I_\mathrm{filt}$ to prevent dust on the fiber face or optics from skewing the S/N down.

We calculate the S/N for each individual image and average them together within each data set to yield a single-rotation S/N. We then coadd images 1--2, 1--3, ..., 1--10 and calculate the S/N for each case. The S/N for images 1--10 is presented as the 10-rotation S/N and each intermediate step as two-rotation S/N, three-rotation S/N, etc.

Far-field images are taken for each data set and analyzed using the maximum intensity rather than the median intensity as the numerator in Equation (\ref{eq:snr}). The far-field speckle pattern is of interest in precision RV spectroscopy, as it is what illuminates the optics and thus affects the line spread function of the instrument. Also, the optical far field and near field are not directly correlated, meaning any result in the near field does not automatically extend to the far field. That being said, we found that all results listed in the following section for the near field are identical to those found when using the far field. Also, mapping the far-field speckle pattern to RV error would require numerical simulations with optical design software and is beyond the scope of this paper. Therefore, we omit the far-field data for conciseness.

\section{Results}
\label{sec:results}

\subsection{Method of Agitation and Fiber Geometry}
\label{subsec:ag_snr}

\begin{figure}
\centering
	\includegraphics[width=\columnwidth]{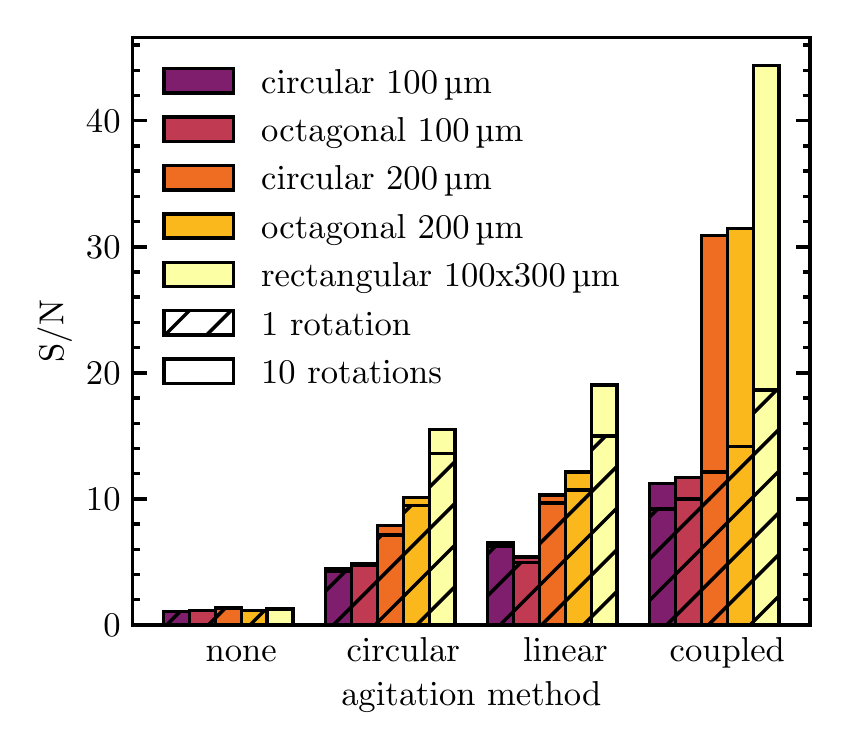}
	\caption{S/N comparison for varying fiber geometries and large-amplitude agitation methods. The S/N is presented for both single-rotation and 10-rotation images.}
\label{fig:ag_snr}
\end{figure}

\begin{figure}
\centering
	\includegraphics[width=\columnwidth]{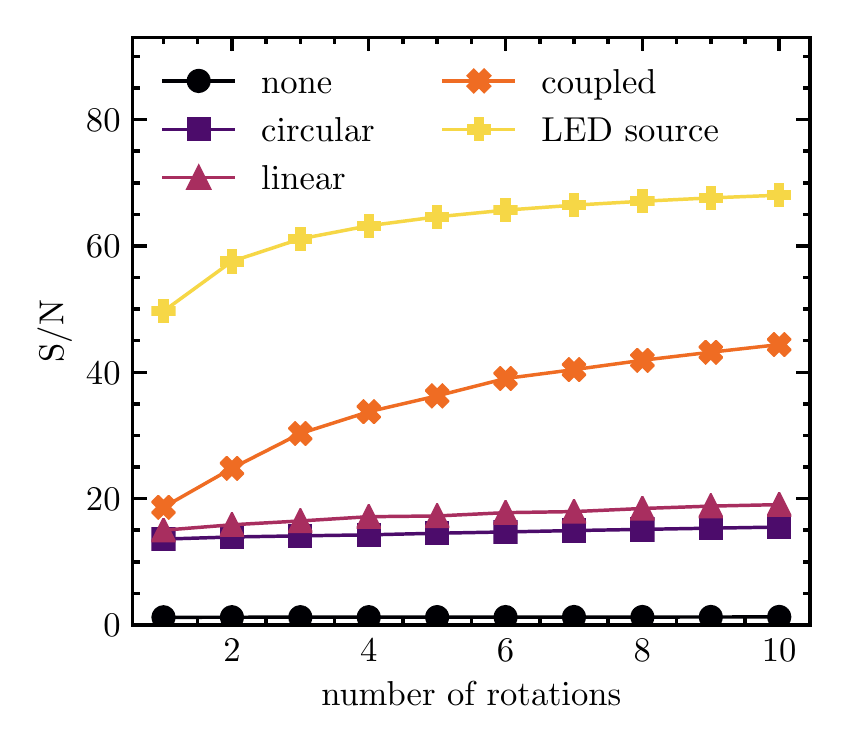}
	\caption{S/N dependence on number of rotations for the \SI{100x300}{\micro\meter} rectangular fiber using various agitation methods. We find similar relationships for the all of the remaining fiber shapes and sizes.}
\label{fig:rect_snr_vs_time}
\end{figure}

We compare the two individual agitation methods and coupled agitation using all of the fibers listed in Table \ref{table:fibers}. The linear agitator is set to an amplitude of \SI{80}{\milli\meter} and the frequency of both agitators to \SI{1.0}{\hertz}. All images are taken with \SI{1.0}{\second} exposures.

Results for the single-rotation and 10-rotation cases are shown in Figure \ref{fig:ag_snr}. Across all fiber shapes and sizes and number of rotations, the linear agitator appears to be slightly better than the circular agitator. There is a similar increase in S/N when looking at coupled agitation in single-rotations. However, coupling the agitation over 10-rotations significantly increases the S/N for all fiber configurations.

To better understand why coupled agitation is more effective at reducing modal noise at longer exposures, we also analyze the effect of each agitation method over multiple rotations, as shown in Figure \ref{fig:rect_snr_vs_time} for the \SI{100x300}{\micro\meter} rectangular fiber. The S/N shows significant improvement beyond a single rotation when coupling the agitation methods. The rate of this improvement is about the same (if not slightly better) than that when the fiber is lit by an LED, a relatively incoherent source. Circular and linear agitation, on the other hand, effectively plateau after the first rotation. We see identical effects for all of the remaining fiber shapes and sizes.

10-rotation images of the three larger fibers using the coupled agitation method are shown in Figure \ref{fig:fiber_improved}. Compared to those shown in Figure \ref{fig:fiber_example}, the speckle patterns that appear when using coupled agitation are nearly nonexistent.

\subsection{Amplitude and Frequency of Agitation}
\label{subsec:amp_freq}

\begin{figure*}[t]
\centering
	\includegraphics[width=\textwidth]{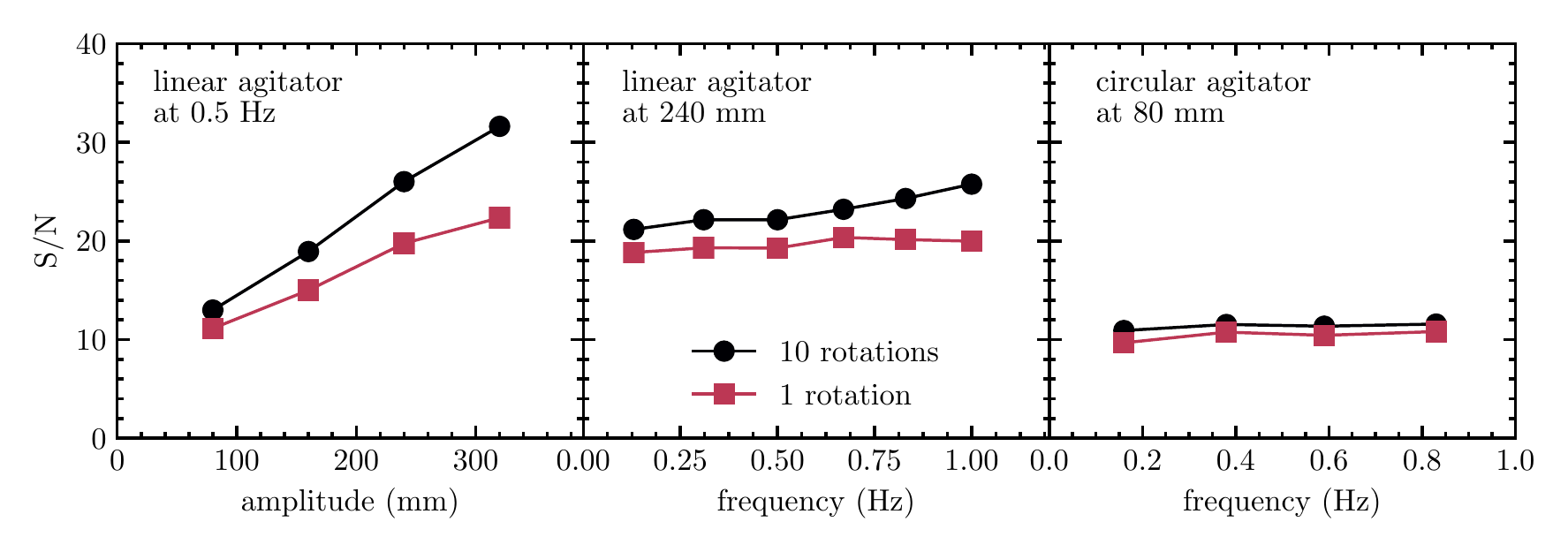}
	\caption{S/N comparison for varying amplitudes using the linear agitator (left) and varying frequencies using each of the linear (center) and circular (right) agitators.}
\label{fig:amp_freq_snr}
\end{figure*}

We use the \SI{100x300}{\micro\meter} rectangular fiber and the two agitators separately to test the effects of agitation amplitude and frequency of rotation on the S/N. We can only test amplitude on the linear agitator and take an image set for each position on the rotating arm. We test frequency on each of the linear and circular agitators at approximately equally spaced frequencies across their entire frequency range.

Results from these tests are shown in Figure \ref{fig:amp_freq_snr}. There is a strong positive correlation between linear agitation amplitude using both the single-rotation and 10-rotation analyses. There also appears to be a slight increase in S/N for the linear agitator at higher frequencies after ten rotations; however, there is no such increase for the single-rotations or any of the frequencies when using the circular agitator.

\begin{figure}
\centering
	\includegraphics[width=\columnwidth]{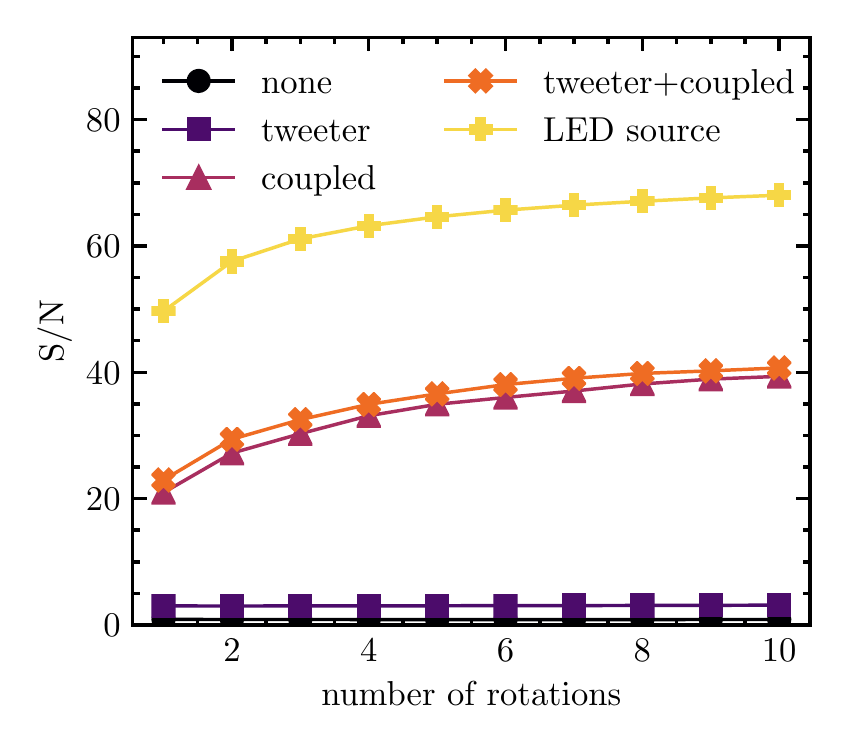}
	\caption{S/N comparison when adding a tweeter to the coupled agitation method for the \SI{100x300}{\micro\meter} rectangular fiber.}
\label{fig:tweeter_snr}
\end{figure}

We also test the high-frequency, low-amplitude tweeter proposed by \citet{Plavchan2013} in tandem with our coupled agitation to see if it supplies significant additional improvement to S/N. We again exclusively use the \SI{100x300}{\micro\meter} rectangular fiber for this test. We set the linear agitator to \SI{240}{\milli\meter} and \SI{0.5}{\hertz} and the circular agitator to approximately the same frequency. The tweeter is set to \SI{20}{\hertz} which has an intrinsic amplitude of \SI{5}{\milli\meter}. All exposure times are set to \SI{2.0}{\second} to match the rotation periods of the large-amplitude agitators. Therefore, the tweeter oscillates 40 times per exposure.

The results are shown in Figure \ref{fig:tweeter_snr}. The tweeter slightly improves S/N regardless of exposure time when compared to the unagitated fiber and when added to coupled agitation. However, the magnitude of this improvement is minimal and is far outweighed by the improvement due to coupled agitation. Also, this improvement to S/N does not compound over time, but rather adds a constant S/N to the coupled agitation regardless of the combined exposure time.

\subsection{Fiber Coupling}

\begin{table}
\centering
\caption{Fiber assemblies tested for the fiber coupling experiment}
	\begin{tabular}{ccc}
	\hline
	Test & First Fiber & Second Fiber \\
	\hline \hline
	1 & Circular \SI{200}{\micro\meter} & Circular \SI{200}{\micro\meter} \\
	\hline
	2 & Circular \SI{100}{\micro\meter} & Circular \SI{200}{\micro\meter} \\
	\hline
	3 & Octagonal \SI{200}{\micro\meter} & Circular \SI{200}{\micro\meter} \\
	\hline
	\end{tabular}
\label{table:fiber_coupling}
\end{table}

It is not uncommon for RV spectrographs to have multiple fiber links for carrying calibration and/or science light from the source (lamp or telescope respectively) to the spectrograph and ultimately the detector. This results in having to couple light from one fiber to another and begs the question, is there a preferred fiber to agitate or must we agitate as many as possible? Agitating the first fiber in a multi-fiber system serves to vary the input illumination of subsequent fibers, similar to the methods proposed by \citet{Mahadevan2014} and \citet{Halverson2014}. However, agitating subsequent fibers may still be necessary to prevent modal noise from re-emerging.

To study the effects of such an architecture, we agitate individual fibers in a multi-fiber assembly where each fiber could have different core sizes and shapes. We test three distinct cases outlined in Table \ref{table:fiber_coupling} and compare them against agitating a single \SI{200}{\micro\meter} circular fiber. For this test, fibers are coupled using one-to-one reimaging optics and fibers are agitated using the linear agitator at \SI{80}{\milli\meter} and \SI{1.0}{\hertz}. When both fibers are agitated simultaneously, they are attached at the same place on the linear agitator, meaning the phase of agitation between the two fibers is constant unlike the previous ``coupled'' agitation tests.

\begin{figure}
\centering
	\includegraphics[width=\columnwidth]{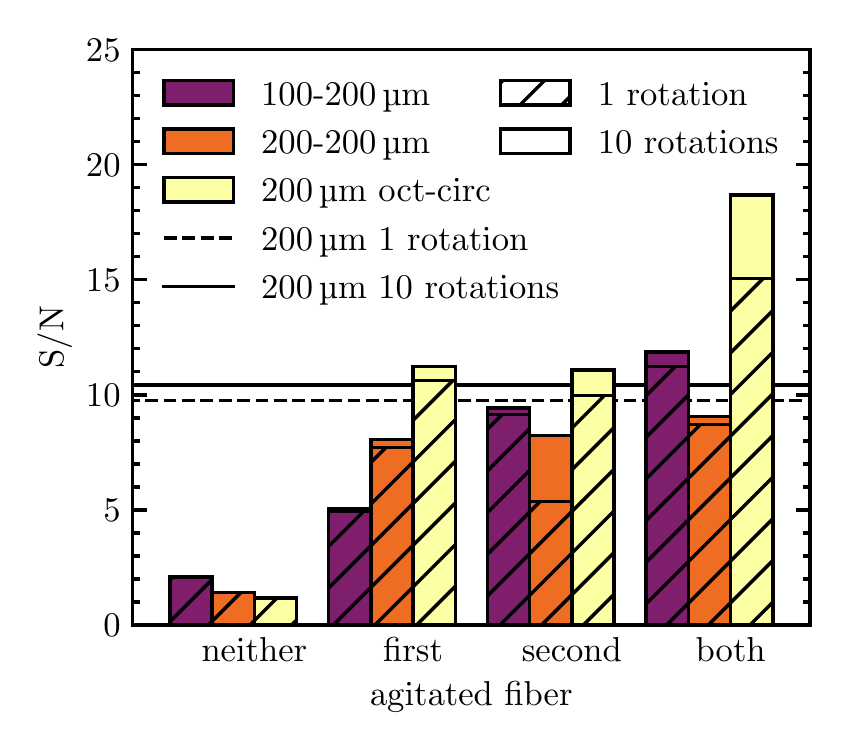}
	\caption{S/N for various arrangements of coupled fibers with varying core diameter and cross-sectional shape. All fibers can be assumed circular unless otherwise stated. The S/N for a single \SI{200}{\micro\meter} circular fiber is also presented for both the singe-rotation (dashed) and 10-rotation (solid) images.}
\label{fig:coupled_fibers}
\end{figure}

The results from these tests are shown in Figure \ref{fig:coupled_fibers}. For the most part, the S/N for each test hovers around 10, the same level at which the S/N would be for an uncoupled \SI{200}{\micro\meter} fiber. However, there are three trends to notice:
\begin{enumerate}
\item When coupling the \SI{200}{\micro\meter} circular fiber to another \SI{200}{\micro\meter} circular fiber, location of agitation does not appear to matter, even if agitating in multiple places.
\item Agitating only the \SI{100}{\micro\meter} circular fiber alone is worse than agitating the \SI{200}{\micro\meter} alone. However, the S/N improves when agitating both of them together.
\item Agitating the \SI{200}{\micro\meter} octagonal fiber or \SI{200}{\micro\meter} circular fiber individually shows no difference in S/N, but agitating both of them significantly improves S/N.
\end{enumerate}
We note that the S/N for single-rotation images when agitating the second fiber in the 200-\SI{200}{\micro\meter} test seems to be abnormally low, but this is alleviated after 10 rotations.

\subsection{Discussion}
\label{subsec:discussion}

Our results can be summarized as follows: the highest S/N is attained when a fiber has been put through as many physical configurations as possible over the length of an exposure. This is best accomplished using a coupled agitation setup comprised of, in our case, linear and circular motion with the highest amplitude possible on each. Due to the changing phase between the two agitators, the resulting total agitation is effectively chaotic. One could conceive a single-element, random agitator as another possible implementation. Note that the motion needs to be continuous to avoid build-up of a static speckle pattern within the exposure.

This conclusion follows from some of the assumptions addressed in previous studies and introduced in Section \ref{subsec:mitigation}. \citet{Grupp2003} actually recommends a chaotic high-amplitude agitator in his statistical study of modal noise. More recently, \citet{Lemke2011}, \citet{McCoy2012}, \citet{Mahadevan2014}, and \citet{Roy2014} find that hand-agitation is consistently better than any form of mechanical agitation. Human motions are inherently less deterministic than mechanical devices, thus resulting in a more chaotic motion. Our combined linear/circular agitator with slightly different oscillation frequencies mimics this behavior since it chaotically reaches many fiber configurations.

Our results also strongly indicate that larger amplitude is much more crucial to mitigating modal noise than increased frequency, as clearly shown in Figure \ref{fig:amp_freq_snr}. We continue to assert that any periodic rotation used as fiber agitation should complete its cycle within a single detector exposure as originally stated by \citet{Baudrand2001}. However, for single-element agitation, increasing the frequency does not show much of a discernible effect per rotation. We do note that the linear agitator does show a slight positive trend with frequency, but we believe this is caused by small random motions in the fiber further from the agitator that are indirectly intensified by the rapid motions of agitation.

Our tweeter tests also help show the importance of amplitude. Even though the high-frequency device is able to place the fiber into many positions over a single exposure, the difference in these configurations is relatively small. Therefore, the speckle pattern is only ``fuzzed'' rather than averaged over the entire fiber face. Adding a tweeter to a large-amplitude agitator does show some small improvements (since extra ``fuzzing'' would naturally increase S/N), but these improvements are significantly overshadowed by simply having large-amplitude chaotic motion.

We are also able to confirm previous results that show better mitigation of modal noise for fibers with larger cross-sectional areas and less azimuthal symmetry. As seen in Figure \ref{fig:ag_snr}, across all agitation methods, the \SI{200}{\micro\meter} fibers fare better than the \SI{100}{\micro\meter} fibers and the rectangular fiber was consistently far better than the others. Our derived expression for the number of modes (Equation (\ref{eq:mode_area})) accurately describes the ratio between each measured S/N shown in Figure \ref{fig:ag_snr}. For example, Equation (\ref{eq:mode_area}) predicts that doubling the diameter of a fiber should double the S/N, which is precisely reflected in Figure \ref{fig:ag_snr}. We find that the rectangular fiber has approximately $\sqrt{2}$ times the S/N of the circular and octagonal fibers across all agitation methods since $C_\mathrm{rect} \approx 2C_\mathrm{circ}$. Although there is no exact trend, the octagonal fibers tend to have a higher S/N than the circular fibers leading us to believe $C_\mathrm{oct}$ is slightly larger than $C_\mathrm{circ}$.

It follows that the location of agitation in a fiber architecture should be on the fiber that propagates the most modes. We verify this as shown in Figure \ref{fig:coupled_fibers}. When coupling two fibers together that propagate the same number of modes, in our case a \SI{200}{\micro\meter} circular fiber with itself and with a \SI{200}{\micro\meter} octagonal fiber, there is no discernible difference in S/N when agitating one over the other. S/N is significantly worsened, however, when a smaller \SI{100}{\micro\meter} circular fiber is agitated instead. Agitating both fibers appears to combine the improvements to S/N caused by agitating each fiber individually, but only when the two fibers have different size or geometry. When the two fibers are identical, S/N is unaffected.

Moreover, we can infer that the location of agitation along a single fiber does not affect S/N. Coupling two \SI{200}{\micro\meter} fibers with a 1:1 ratio is effectively adding their lengths together and creating a single fiber. As shown by Figure \ref{fig:coupled_fibers}, the location of agitation is irrelevant for such a situation, especially for 10-rotation exposures. Therefore, the agitator could be placed anywhere along the length of the fiber (preferably far away from the spectrograph) and it will produce the same magnitude effect on modal noise. This conclusion relies on only one test, however, so it will require further study to absolutely confirm.

Coupled fiber test cases with light loss due to improperly matched \'etendue, such as coupling light from a \SI{200}{\micro\meter} fiber into a \SI{100}{\micro\meter} fiber, were not covered for this paper and will require further study. However, we suspect that the best modal noise mitigation will consistently occur when agitating the fiber that propagates the most modes, since improvements to modal noise S/N occur across the entire near field and far-field projections and should be unaffected by truncation.

\section{RV Precision}
\label{sec:rv_precision}

\begin{figure*}
\centering
	\includegraphics[width=\textwidth]{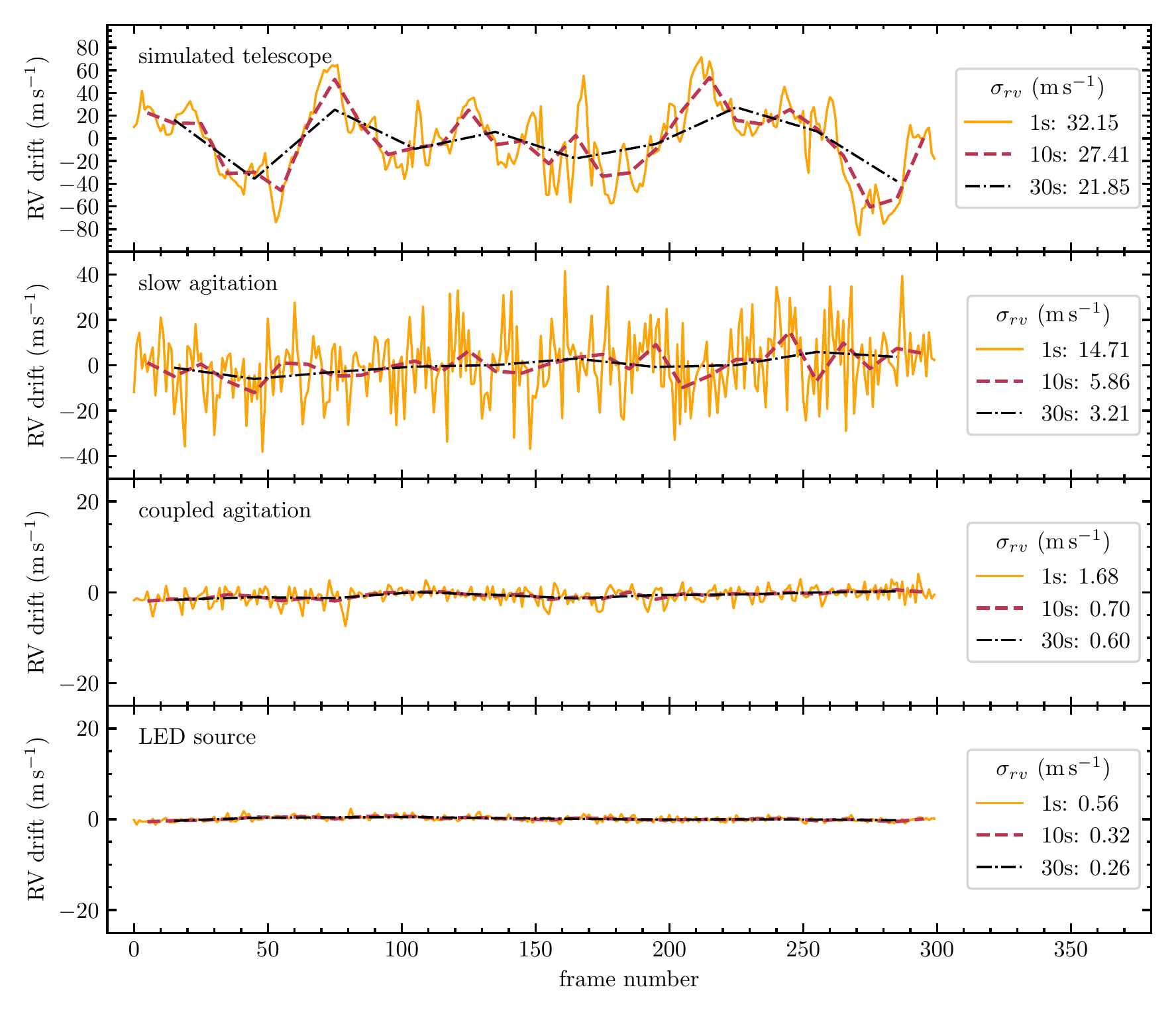}
	\caption{Centroid drift and resultant RV error for a fiber moved by a simulated telescope (first), slowly agitated fiber (second), coupled agitated fiber (third), and LED illumination (fourth). Each line represents a different exposure time each with their own calculated RV error. Note that the scale for the top three plots are different.}
\label{fig:rv_error}
\end{figure*}

As discussed in Section \ref{sec:modal_noise_intro}, optical fiber modal noise is an issue of centroid drift as well as diminished S/N. To observe how the centroid actually drifts over time, we test our agitation method on the \SI{100x300}{\micro\meter} rectangular fiber over three hundred \SI{1.0}{\second} exposures.

The resultant RV precision ($\sigma_{RV}$) due to a shifting speckle pattern centroid at the end of a fiber is calculated as
\begin{equation}
\sigma_{RV} \approx \frac{c}{R} \frac{\sigma_d}{D},
\label{eq:rv_error}
\end{equation}
where $c$ is the speed of light in a vacuum, $R$ is the resolution of the spectrograph, $\sigma_d$ is the standard deviation of fiber near-field centroid drift in the dispersion direction, and $D$ is the slit width (or short-end length of a rectangular fiber). Notice that for a $R=150,000$ spectrograph fed by a \SI{33}{\micro\meter} slit attempting to reach \SI{1.0}{\meter\per\second} RV precision per line, the required stability of the centroid along the dispersion direction is \SI{0.0165}{\micro\meter}.

Importantly, $\sigma_{RV}$ is only the RV error \textit{per resolution element} or \textit{per line} from a wavelength calibration source. Averaging over $N$ lines with independent modal structure, we can divide $\sigma_{RV}$ by $\sqrt{N}$ to approximate total RV error. $N$ may not necessarily equal the total number calibration lines, however, since two neighboring wavelengths may propagate with the same number of modes and thus the same modal structure. Recall from Equation (\ref{eq:mode_area}) that the number of supported modes is a function of wavelength. With a high number (thousands) of propagating modes, we do not expect adding or subtracting a single mode will result in a statistically independent modal noise structure. Though, such a study is beyond the scope of this paper.

However, assuming a difference in 10 supported modes makes the structure effectively independent, we assert
\begin{equation}
N \approx \frac{M_s(\lambda_{min}) - M_s(\lambda_{max})}{10},
\label{eq:N}
\end{equation}
where $\lambda_{min}$ and $\lambda_{max}$ are, respectively, the minimum and maximum wavelengths of the relevant calibration region. Note that the calibration source needs to be sufficiently dense (i.e. the number of lines is greater than $N$) to properly use this approximation. Even though Equation (\ref{eq:N}) has not been empirically tested, since it requires a comprehensive study on the systematic correlation of modal noise, we believe it to be a rather conservative estimate for statistical reduction.

We derive Equation (\ref{eq:rv_error}) from the low velocity approximation of the relativistic Doppler effect
\begin{equation}
\frac{\Delta \lambda}{\lambda} = \sqrt{\frac{1 + v/c}{1-v/c}} - 1 \approx \frac{v}{c},
\label{eq:doppler_effect}
\end{equation}
where $\Delta \lambda$ is the measured shift in wavelength at wavelength $\lambda$ on the spectrograph for a star moving at velocity $v$ relative to Earth. The resolution of a spectrograph is
\begin{equation}
R = \frac{\lambda}{\Delta \lambda_R},
\label{eq:resolution}
\end{equation}
where $\Delta \lambda_R$ is the width of the spectrograph resolution element in terms of wavelength bandwidth. Centroid shifts in the near field of the fiber face can thus be equated to a measured wavelength shift at the focal plane of the spectrograph:
\begin{equation}
\frac{\Delta d}{D} = \frac{\Delta \lambda}{\Delta \lambda_R}.
\label{eq:spx_shift}
\end{equation}

Combining Equations (\ref{eq:doppler_effect})--(\ref{eq:spx_shift}) we show that
\begin{equation}
\frac{v}{c} \approx \frac{\Delta \lambda}{\lambda} = \frac{1}{R} \frac{\Delta \lambda}{\Delta \lambda_R} = \frac{1}{R} \frac{\Delta d}{D}.
\end{equation}
If we take the standard deviation of the data from each side of this equation ($v \rightarrow \sigma_{RV}, \Delta d \rightarrow \sigma_d$) and move $c$ to the right side, we get Equation (\ref{eq:rv_error}).

The idealized agitation method we use to test RV precision includes the circular agitator oscillating at \SI{1.1}{\hertz} and the linear agitator set at \SI{240}{\milli\meter} and \SI{1.0}{\hertz}. Keeping the two agitators at slightly different frequencies means that phase between them constantly changes and a large range of fiber configurations are reached after about \SI{10}{\second}.

We compare this idealized method to the LED source (low modal noise), a slowly agitated fiber (high modal noise), and a fiber moved as if it were attached to a telescope (very high modal noise). For the slow agitation test, we set only the linear agitator at \SI{80}{\milli\meter} and \SI{0.03}{\hertz} meant to simulate a worst-case simple harmonic agitation method. To simulate the telescope motions, we use the linear agitator with \SI{80}{\milli\meter} amplitude but do not rotate continuously. Rather, we leave the agitator off most of the time except to turn it on briefly (about a quarter or half rotation) every 50 images. This approximates two conditions of the telescope: nearly still while tracking a star and sudden motion while switching targets.

To calculate the RV error, we first find the centroid within the fiber face for each image, relative to the center of the fiber in order to remove the observable ($\sim 2$ pixel) drift of the fiber stage. We then project this centroid drift along the short axis of the rectangle. We also average these centroids over sets of 10 and 30 images to approximate the centroid drift for longer exposure times. For each length of exposure, we calculate the standard deviation of the centroid drift and convert all values to RV using Equation (\ref{eq:rv_error}). The dispersion direction for a rectangular fiber is along the short end, meaning that the slit width $D$ is \SI{100}{\micro\meter}. The resolution of EXPRES is 150,000 for a \SI{33x132}{\micro\meter} rectangular fiber, so we use $R=50,000$ in Equation (\ref{eq:rv_error}) since our test fiber is three times as wide as the EXPRES fiber in the dispersion direction.

The results, shown in Figure \ref{fig:rv_error}, indicate that continuous agitation is essential to control modal noise. The simulated telescope yields RV errors above \SI{20}{\meter\per\second} that do not average out well with longer exposures. Using our idealized method of coupled agitation reduces errors from slow agitation by about 5--8 times and are so far minimized to about \SI{60}{\centi\meter\per\second} when using \SI{30}{\second} exposures. The minimum RV error we could measure in this test, by using the LED source, was \SI{26}{\centi\meter\per\second}. Therefore, our coupled agitation method is less than a factor of 3 from reaching our noise floor.

The calculated \SI{60}{\centi\meter\per\second} error for coupled agitation is the RV error \textit{per line} in the spectrograph. Thus, the total RV error could be reduced to below \SI{10}{\centi\meter\per\second} with only 36 mode-independent calibration lines. EXPRES is using a laser frequency comb with approximately {\SI{14}{\giga\hertz}} line spacing across 450-{\SI{700}{\nano\meter}} fed by a {\SI{33x132}{\micro\meter}} rectangular fiber at $f/3.0$ resulting in almost 17,000 calibration lines. Applying Equation (\ref{eq:N}), EXPRES will have $N=350$, reducing the expected RV error of the instrument to less than {\SI{3.2}{\centi\meter\per\second}}.

\section{Agitation and FRD}
\label{sec:frd}

There is concern in the RV community that mechanical agitation will exacerbate FRD within fiber architectures. FRD is defined as when the focal ratio of the output of an optical fiber is less than that of the injected light. In RV spectroscopy, this means that the focal ratio of the telescope may not be preserved when transmitting light to the spectrograph and losses due to FRD-induced vignetting could occur.

FRD can be worsened through mechanical deformation \citet{Ramsey1988}, classified as macrobending (changes to the radius of curvature of bends in the optical fiber) or microbending (deformations smaller than the diameter of the fiber). These bends can couple light into previously unrealized modes thus causing dispersion in the fiber's far field. \citet{Powell1984}, \citet{Engelsrath1986}, and \citet{Ramsey1988} find that macrobending on its own has little effect on FRD. The fear is rather that violently macrobending the fiber (such as through agitation) may stress the fiber and cause microbending, a more severe detriment to FRD.

Although it has been shown by \citet{Sablowski2016} that their agitator (which has similar frequency and amplitude to our own) has little effect on FRD, we thought it wise to similarly test FRD using our agitator. We use the {\SI{100x300}{\micro\meter}} rectangular fiber and coupled agitation at approximately {\SI{0.5}{\hertz}} for this test. Using the FCS, we image the far field at various injected focal ratios ($f/3.0$, $f/4.0$, $f/5.0$) selected with an iris on the input pupil. We then determine the output focal ratio by calculating the Gaussian beam diameter of the far-field image (where the edges are set at 1/${e^2}$ of the maximum amplitude) and compare this value to images taken with the output pupil iris set to known diameters.

\begin{figure}
\centering
	\includegraphics[width=\columnwidth]{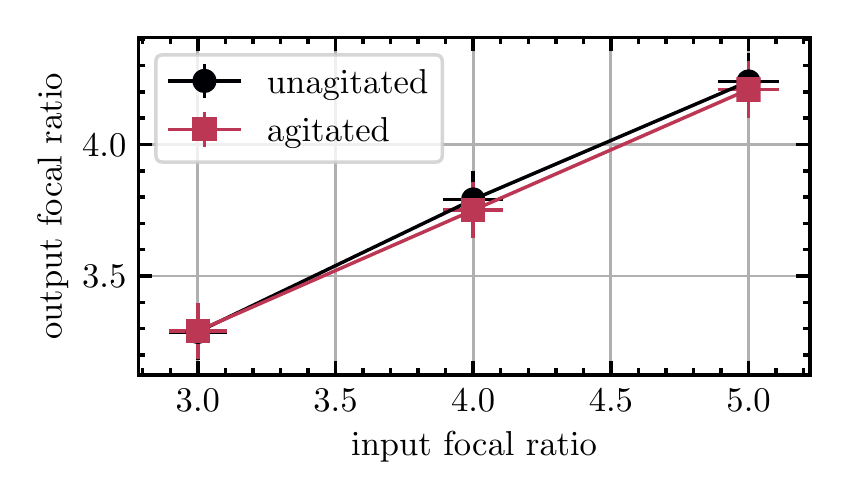}
	\caption{Relationship of the input and output focal ratios for the {\SI{100x300}{\micro\meter}} rectangular fiber both with and without agitation.}
\label{fig:frd}
\end{figure}

The results for this test are shown in Figure {\ref{fig:frd}}. Across all three injected focal ratios, the output focal ratio is decreased by less than 0.1 when the fiber is agitated. Thus, there is apparently minimal effect to FRD caused by agitating the fiber with coupled agitation, our most stressful scheme. This does not account for long-term wear-and-tear effects of high-amplitude agitation on FRD, a topic that requires further study, but does show that our coupled agitation method is gentle enough to not cause any immediate devastating issues.

\section{Summary and Application}
\label{sec:conclusions}

\begin{figure}
\centering
	\includegraphics[width=\columnwidth]{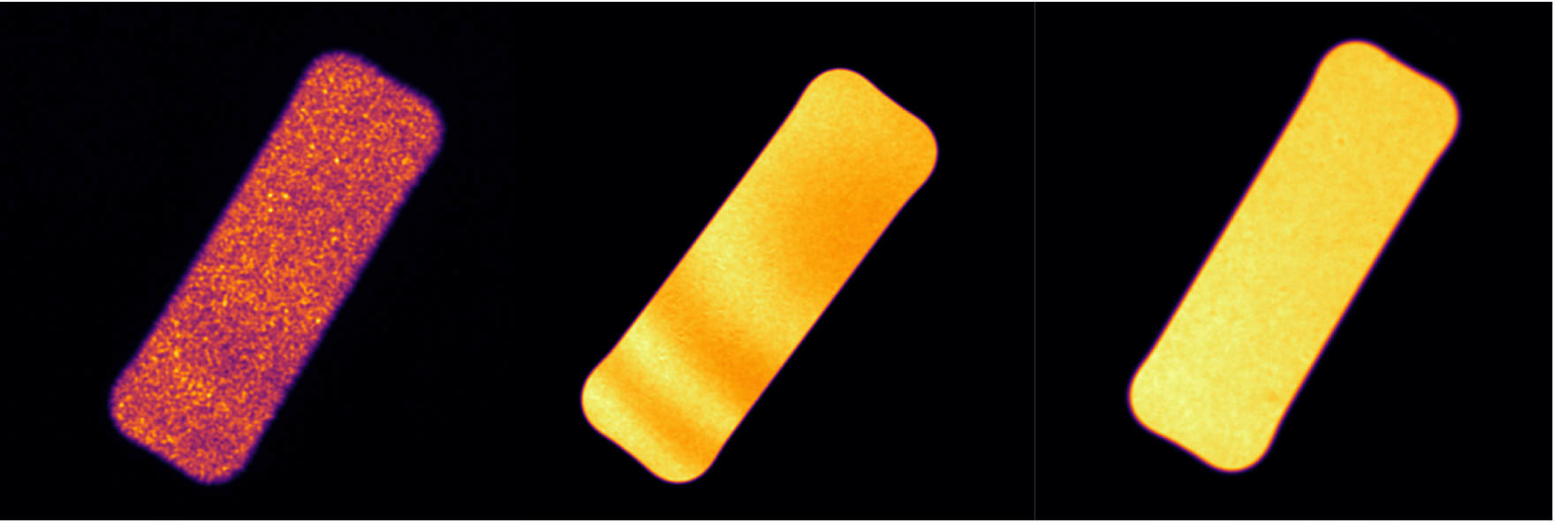}
	\caption{Comparison of the long-term agitation methods used in Section \ref{sec:rv_precision}---slow agitation (left), coupled agitation (middle), LED source (right)---as \SI{10}{\second} exposures. Brightness directly scales with photon count in these images.}
\label{fig:fiber_rv_error}
\end{figure}

\begin{figure}
\centering
	\includegraphics[width=\columnwidth]{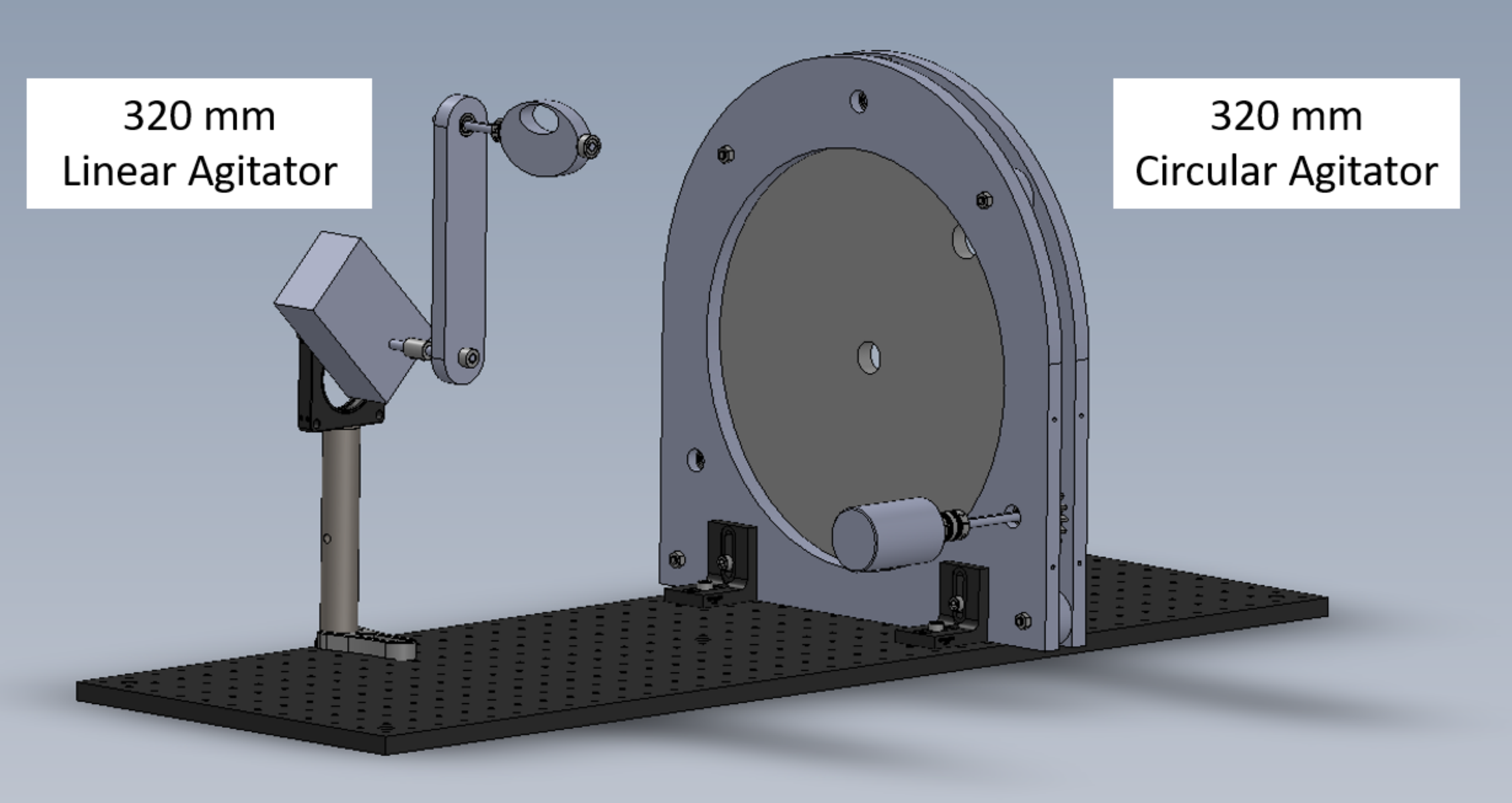}
	\caption{Rendering of the proposed fiber agitator for EXPRES. Both the linear and circular agitators have independent DC motors operated by remote computer control. This agitator will sit on a shelf across the room from the spectograph and will agitate all fibers that enter the spectrograph during both stellar and calibration exposures.}
\label{fig:agitator_model}
\end{figure}

We have tested a wide swath of agitation parameter space with the goal of further understanding the mechanisms behind fiber agitation as a method for modal noise mitigation. Our conclusions, as an update to the previous assumptions introduced in Section \ref{subsec:mitigation}, are as follows:
\begin{enumerate}
\item Agitating the fiber with the \textbf{most modes}, regardless of its placement in the fiber architecture, yields the best S/N. This typically means the fiber with the largest core size and lowest azimuthal symmetry.
\item \textbf{Large-amplitude} agitation is much more important than high-frequency agitation as long as the agitation method reaches at least one full rotation within an exposure. Adding a ``tweeter'' shows minimal improvement over large-amplitude motion alone.
\item \textbf{Chaotic agitation}, such as the motion created by our ``coupled'' agitator, is much more effective at mitigating modal noise than typical harmonic agitation and continues to improve S/N over multiple rotations. The frequency of this method should place the fiber into as many configurations as possible over a single exposure without over-stressing the fiber.
\item \textbf{Agitating more fibers} in a fiber architecture, especially when the fibers have different geometry (size/shape), will help increase S/N. Using a single agitator with one fiber looped over it several times, however, helps only slightly. Rather, adding a second independent agitator (i.e. chaotic agitation) increases S/N significantly more.
\end{enumerate}

As shown in Figure \ref{fig:fiber_rv_error}, our two high-amplitude agitators oscillating with varying phase reduce modal noise to levels almost indiscernible from a fiber propagating broadband light after only 10 cycles over an exposure. Since agitation hardly affects throughput efficiency, chaotic fiber agitation can be used with relatively dim light sources, allowing for direct application to the next-generation of precision RV spectrographs.

It is important to note that there has not yet been a long-term study on how shaking an optical fiber may affect attenuation or the potentiality to completely break the fiber. We demonstrate that FRD is unaffected in the short-term and, so far, our fibers have not shown detrimental effects due to the aforementioned agitation methods. Our results also show that high-amplitude agitation (i.e. macrobending) in multiple places is more advantageous than quick motions in one location, thus, we believe that microbending---a primary cause of focal ratio degradation---can be more easily avoided. However, any project that wishes to mechanically agitate fibers should take care to avoid over-bending or stretching their fibers and shake them with a minimal amount of aggression.

As part of the EXPRES fiber architecture, we will be employing the quasi-chaotic agitation technique detailed in this paper. We will be combining a \SI{320}{\milli\meter} amplitude circular and linear agitator similar to those seen in Figure \ref{fig:agitators}, but with greater stability to support an entire reinforced wrap of cables. A rendering of this design is shown in Figure \ref{fig:agitator_model}. The two agitators will oscillate at slightly different frequencies (e.g., 0.9 and \SI{1.0}{\hertz} for a wide range of phases over a \SI{10}{\second} exposure) at the maximum speed deemed safe for the fibers. This device will agitate all of the fibers that immediately enter the spectrograph, which includes, most importantly, our rectangular fiber.

We recommend that other precision RV spectrographs consider the results found in this paper when designing their own fiber agitators. Since it only affects the fibers between light sources and the spectrograph, such improved agitation methods can even be added to previously commissioned spectrographs to increase S/N and reduce potential false positives. We would recommend simply adding a second independent agitator anywhere along the fiber train to help induce more chaoticism. Modal noise is not a problem that should be treated lightly, as its mitigation will help usher in the next-generation of RV spectroscopy and aid in the search for Earth-sized worlds.

\acknowledgments

We would like to acknowledge NSF Major Research Instrumentation Award AST 1429365, as well as an NSF Advanced Technologies Instrumentation Award AST 1509436. The author would also like to acknowledge Gabor Furesz for assistance with the Fiber Characterization Station design, Saki Kamon and Kristoffer Acu\~na for their data-taking contributions to this project, and the anonymous referee for their thorough comments. The Fiber Characterization Station was built with support from the Fund for Astrophysical Research, Inc. This material is based upon work supported by the National Science Foundation Graduate Research Fellowship under Grant No. 2017242370.

\bibliographystyle{aasjournal}
\bibliography{modal_noise_mitigation}

\end{document}